\documentclass[12pt,preprint]{aastex} 

\tighten 

\slugcomment{Accepted for publication in {\it the Astrophysical Journal Letters
2002 August 22}} 
\def\fdg{\hbox{$^\circ\!.$}}
\def\lsim{\lower.5ex\hbox{$\; \buildrel < \over \sim \;$}} 
\def\gsim{\lower.5ex\hbox{$\; \buildrel > \over \sim \;$}} 
\def\lax {\ifmmode{_<\atop^{\sim}}\else{${_<\atop^{\sim}}$}\fi} 
\def\gax {\ifmmode{_>\atop^{\sim}}\else{${_>\atop^{\sim}}$}\fi} 
 
\def\gtorder{\mathrel{\raise.3ex\hbox{$>$}\mkern-14mu 
\lower0.6ex\hbox{$\sim$}}} 
\def\ltorder{\mathrel{\raise.3ex\hbox{$<$}\mkern-14mu 
\lower0.6ex\hbox{$\sim$}}}

\def\pmb#1{\setbox0=\hbox{#1}%
\kern-0.015em\copy0\kern-\wd0 
\kern0.03em\copy0\kern-\wd0 
\kern-0.015em\raise0.0433em\box0 }

\begin{document} 

\title{Effects of Resonance in Quasiperiodic Oscillators of Neutron Star
Binaries} 

\author{Lev Titarchuk \altaffilmark{1,2}}

\altaffiltext{1}{George Mason University/CEOSR/NRL;
lev@xip.nrl.navy.mil}
\altaffiltext{2}{NASA Goddard Space Flight Center, code 661, 
Laboratory for High Energy
Astrophysics, Greenbelt MD 20771; lev@lheapop.gsfc.nasa.gov}

\shorttitle{Resonance and Damping Effects}
\shortauthors{TITARCHUK}

\begin{abstract} 

Using a large quantity of Rossi X-ray Timing Explorer  data presented in the literature
 I  offer a detailed investigation into the  accuracy of quasiperiodic oscillations
(QPO) frequency determination. The QPO phenomenon seen in X-ray binaries is possibly a result 
of the resonance of the  intrinsic (eigen) oscillations and harmonic driving forces
of the system. 
I show that the resonances, in the presence of the  damping of oscillations, occur
at the frequencies which are systematically and randomly shifted with respect to
the eigenfrequencies of the system. The shift value strongly depends 
on the damping rate which is measured by the halfwidth of the QPO feature.  
Taking into account this effect I analyze the QPO data 
for four  Z-sources: Sco X-1, GX 340+0, GX 5-1, GX 17+2  and two atoll sources:   
4U 1728-34, 4U 0614+09. 
The transition layer model (TLM)  predicts the existence of the invariant quantity: 
$\delta$, an inclination angle 
of the magnetospheric axis with respect to the normal to the disk.  
I calculate $\delta$ and  the error bars of $\delta$ using the resonance shift and 
I find that the inferred  $\delta-$values are consistent with  constants for these
four  Z-sources,  
where horizontal branch oscillation and  kilohertz 
frequencies have been detected and correctly  identified.
It is shown that the inferred $\delta$ are in the range between 5.5 and  
6.5 degrees. 
I conclude that the TLM seems to be compatible with data.

\end{abstract} 


\section{Introduction} 

Kilohertz quasi-periodic oscillations (QPOs) have been discovered by
the Rossi X-ray Timing Explorer (RXTE) in a number of low mass X-ray
binaries (Strohmayer et al.  1996, van der Klis et al. 1996). The
presence of two observed peaks with frequencies $\nu_1$ and $\nu_2$ in
the upper part of the power spectrum became a natural starting point
in modeling the phenomena.  Attempts have been made to relate $\nu_1$
and $\nu_2$ and the peak separation $\Delta \nu= \nu_2-\nu_1$ with the
neutron star (NS) spin. In the sonic point beat frequency (SPBF) model  by Miller, Lamb \&
Psaltis (1998) the kHz peak separation $\Delta \nu$ is considered to be
close to the NS spin frequency and thus $\Delta \nu$ is predicted to
be constant [see also the updated version of the SPBF (Lamb \& Miller 2001) 
in which $\Delta \nu$ is allowed to be variable].
Observations
of kHz QPOs in a number of binaries (Sco X-1, 4U 1728-34, 4U 1608-52,
4U 1702-429 and etc) show that the peak separation decreases
systematically when kHz frequencies increases (see a review by van der
Klis 2000).  

There are two other  models in the literature which infer the
relations between $\nu_1$, $\nu_2$ and $\nu_{HBO}$.
The relativistic precession (RP) model  involves  high speed
particle motion in strong gravitational fields, leading to
oscillations of the particle orbits. 
Stella et al. (1999) studied precession of the particle orbit under influence of
a strong gravity due to General Relativity (GR) effects. 
The correlation between high frequency (lower kHz frequency) and low frequency 
(broad noise component) QPOs previously found by Psaltis, Belloni \& van der Klis (1999) for black hole (BH) and 
 neutron star (NS) systems has been recently extended 
 over two orders of magnitude by Mauche (2002) to white dwarf (WD) binaries.
With the assumption that the same mechanism produces the QPO in  
WD, NS and BH binaries, Mauche argues that the data exclude RP, SPBF models (as well as any model requiring either the
presence or absence of a stellar surface or a strong magnetic field).
 
The transition layer model (TLM) was introduced by Titarchuk, Lapidus
\& Muslimov (1998), hereafter TLM98, to explain the dynamical
adjustment of a Keplerian disk to the innermost sub-Keplerian boundary
conditions (e.g. at the NS surface).  
The low branch frequency $\nu_L$, the Kepler frequency $\nu_{\rm K}$
and the hybrid frequency $\nu_h$, as they are introduced by OT99 are
eigenfrequencies of the oscillator.  In the TLM framework  the observed horizontal branch oscillation (HBO) 
frequencies $\nu_{HBO}$,
$\nu_1$ and $\nu_2$ can be interpreted  as the {\it resonance} frequencies near 
$\nu_{L}$, $\nu_{\rm K}$ and $\nu_{h}$  which are broadened  as a result of
the (radiative) damping in the oscillator (see TLM98, Eq. 15).
Furthermore, the resonance frequencies $\nu_{HBO}$, $\nu_1$, $\nu_2$
are shifted with respect to the eigenfrequencies $\nu_L$, $\nu_{\rm
K}$, $\nu_h$.  The frequency shift and random errors of the
eigenfrequencies depend on the damping rate of oscillations $\lambda$
(see details in \S 3).  {\it One should keep in mind that the systematic
and random errors in the centroid frequency determination due to this
resonance shift can be a factor of a few larger than the statistical
error in the determination of the centroid frequency.}  
It is worth noting that an interpretation of the observed QPO phenomena in the framework of
any oscillatory (or wave propagation) model requires the aforementioned corrections due to 
resonance in the presence of damping. For the observed Q-values ($Q=\nu/2\lambda)$ of a few, 
the damped intrinsic  harmonic oscillations decay very quickly (less than $10^{-2}$ s) and thus
only the forced oscillations can exist for time periods of order  $10^{2-3}$ seconds and longer
(for details  see \S 3.1).  There remains the question of what kind of the observational arguments 
can support  the resonance mechanism  for QPO phenomena. I address to this issue in \S 4.     
The goal of this {\it Letter} is to demonstrate how 
the resonance shift  corrections are important for  the accurate interpretation of the 
QPO data.  The {\it Letter} reports the results of the QPO data interpretation
from four Z-sources: Sco X-1, GX 340+0, GX 5-1, GX 17+2  and two atoll sources:   
4U 1728-34, 4U 0614+09 and two atoll sources,  collected by RXTE during 6 years of 
observations.  In \S 2, I present the origins 
of the  RXTE QPO data  used in the
present investigation.   In \S 3 I  provide details of the resonance
effect on the eigenfrequency restoration using the observed QPO
frequencies and in \S 4 I offer comparisons of the predictions of the TLM model with
the RXTE observations. Summary and conclusions are drawn in \S 5. 

\section{Source  and Data Selection}  
I used the QPO data from Sco X-1 [van der Klis et al. (1996), (1997),
Bradshaw et al. (2002)]; from GX 340+0 (Jonker et al. 2000a); 
from GX 5-1 (Jonker  et al. 2002); from GX 17+2 (Homan et al. 2002);
from 4U 0614+09 (van Straaten et al. 2000, 2002) and from 4U 1728-34 
(van Straaten et al. 2002).

\section{Resonance effects in  weakly nonlinear oscillators: relevance to observed QPO
phenomenon in X-ray binaries} 

The observational appearance of asymmetric  Lorentzian QPO features,  
 HBO  and low frequency harmonics  (e.g. van der Klis 2000) is the first indication
 of the classical resonance phenomena  along with a combination frequency effect 
 well established  in weakly nonlinear oscillating systems
(e.g. Landau \& Lifshitz 1965, hereafter LL). Similar effects are described by
radiation theory in the context of  atom excitation (as the classical damped  harmonic 
oscillator) which is driven  by a harmonic  wave (see e.g. Mihalas 1970).   

OT99 formulate the QPO problem in the framework 
of an oscillator in the rotational frame of
reference. Taking the magnetospheric rotation with angular velocity 
$\bf{\Omega=2\pi\nu_{mg}}$ 
(not perpendicular to the plane of the equatorial disk) the small amplitude oscillations of
the fluid element thrown into the magnetosphere are described by 
a three dimensional oscillator (see OT99, Eqs 2-4).
When the angle $\delta$ , between ${\bf \Omega}$  and the vector normal to plane of radial
oscillations is small then 
three dimensional intrinsic oscillations are split into  two independent eigenmodes:
the radial mode with the hybrid eigenfrequency $\nu_h=(\nu_{\rm K}^2+4\nu_{mg}^2)^{1/2}$ and 
the vertical mode with
the low branch eigenfrequency $\nu_L=2\nu_{mg}(\nu_{\rm K}/\nu_h)\sin \delta$.
Thus one can consider two independent harmonic oscillators under influence of a damping
force probably the radiation drag force (proportional to the velocity of 
displacement) as introduced by TLM98. 
The restoring force in these oscillators is only in the first approximation proportional
to the dispacement with the proportionality coefficient equal to the square of the
eigenfrequency. The interplay between the gravitational, Coriolis
forces and pressure gradient maintains the oscillations with frequencies close to 
$\nu_h$  and $\nu_L$ but not exactly at these frequencies. In the next approximation
one needs to introduce the nonlinear terms for the restoring force (see LL and Titarchuk 2002).
Thus the small amplitude oscillations of the fluid element in the magnetosphere can be 
described by two equations
\begin{equation}
\ddot{x}+2\hat{\lambda}_1\dot x+\omega_h^2x =A_1e^{i\omega t}+\alpha_{1,1} x^2 +2\alpha_{1,2} x
z+\alpha_{1,3}z^2,  
 \end{equation}
 \begin{equation}
\ddot{z}+2\hat{\lambda}_2\dot z+\omega_L^2z =A_2e^{i\omega t}+\alpha_{2,1} x^2 +
2\alpha_{2,2} xz+\alpha_{2,3}z^2,  
 \end{equation}
where $x$ and $z$ are the radial and vertical components of the displacement vector
respectively. Here we present equations (2-4) from OT99 (written in Cartesian coordinates) using 
the eigenvector basis (radial and vertical eigenvectors).
$\hat{\lambda}_1$ and $\hat{\lambda}_2$ are the damping rates for x and z components;
$\omega_h=2\pi\nu_h$, $\omega_L=2\pi\nu_L$ are angular hybrid and low branch frequencies
respectively (OT99), 
$\omega$ is a frequency of the harmonic driving force and $A_1$, $A_2$ are amplitudes
of x and z components of the driving force,  $\alpha_{k,l}$, ($k=1,2$, $l=1,2,3$)
are coefficients at the weakly nonlinear terms of the restoring force.  
\subsection{Harmonic oscillations}
The linear terms of the solution of equations  (1-2) are 
\begin{equation}
x=C_{1,1}f_{-}(t,\omega_h, \hat{\lambda}_1)+
C_{1,2}f_{+}(t,\omega_h, \hat{\lambda}_1)
+ B_1(\omega_L)\exp{(i\omega t)},
\end{equation}
\begin{equation}
z=C_{2,1}f_{-}(t,\omega_L, \hat{\lambda}_2)+
C_{2,2}f_{+}(t,\omega_L, \hat{\lambda}_2) 
+ B_2(\omega_h)\exp{(i\omega t)},
\end{equation}
where 
$f_{\pm}(t,\omega_{\ast}, \hat{\lambda})=
\exp{\{-[\hat{\lambda}t\pm it(\omega_{\ast}^2-\hat{\lambda}^2)^{1/2}]\}}$, and
$B_{k}(\omega_{\ast})=A_k/[\omega_{\ast}^2-\omega^2+2i\hat{\lambda}_k\omega]$.
The first two terms in Eqs. (3-4)  represent  the damped harmonic oscillations 
(it is assumed  that $\omega_{\ast}/\hat{\lambda}_k>1$). They  
decay very rapidly within a time interval $1/\hat{\lambda}_{k}$ which is of order
of $<10^{-2}$ s. In fact, a typical measured $\lambda_{k}=\hat{\lambda}_{k}/2\pi$
are always higher than a few hertz (see details below).
Thus only  the forced oscillations can exist for   time periods of order $10^{2-3}$
seconds and longer.  
The absolute values of the forced oscillation  amplitudes   
\begin{equation}
b_k=\frac{A_k}{(2\pi)^2[(\nu_{0k}^2-\nu^2)^2+4\lambda_k^2\nu^2]^{1/2}},~~~~~~k=1,2
\end{equation}
(where  $B_k=b_ke^{i\theta}$, $\nu_{0k}=\nu_h,~\nu_L$ for $k=1,~2$
respectively and $\nu=\omega/2\pi$)
has a maximum at
$\nu_{rk}=(\nu_{0k}^2-2\lambda_k^2)^{1/2}$. 
Below we omit the subscript $k$ for  simplicity of the  presentation 
having in mind the particular case  for the eigenfrequency $\nu_0$. 
Hence the frequency shift of the resonance
frequency $\nu_{r}$ with respect to the eigenfrequencies $\nu_{0}$ 
because of damping is
\begin{equation}
\eta=(\nu_{0}^2-2\lambda^2)^{1/2}-\nu_{0}.
\end{equation}
We find that {\it the maximum of the main resonance amplitude for the linear 
oscillations is not precisely at the eigenfrequency $\nu_0$}, but rather
{\it it is shifted to the frequency $\nu_r=\nu_0+\eta$}.  For a linear
oscillator $\nu_r$ depends on the damping rate $\lambda$.
From Eq. (6) it follows that  
$\eta\approx -\lambda^2/\nu_{0}=-\lambda/2Q$  when $Q=\nu_{0}/2\lambda\gg 1$.
It is easy to show (see also LL for details)  that the half-width of the resonance 
$\Delta\nu_{HW}\approx\lambda$ for $Q\gg 1$.
Taking a differential from the left and  right hand sides of the formula for $\nu_r$  
one can derive a formula for the random error of the eigenfrequency
determination $\Delta \nu_0$ due to the resonance effect:
\begin{equation}
\Delta\nu_{0}=\nu_{r}\Delta\nu_{r}/\nu_{0}+2\lambda\Delta\lambda/\nu_0.  
\end{equation}

\subsection{Anharmonic oscillations and p:q resonance in the QPO sources}
The resonance in weakly nonlinear systems occurs at the
eigenfrequencies of the system $\nu_0$ when the frequency of the
driving force  is $\gamma\approx p\nu_0/q$ and $p$, $q$ are
integers (LL). The observed harmonics and subharmonics of HBO frequencies (Jonker et al. 
2000a, 2000b 2002) are probably results of the resonance effect in the weakly nonlinear system. 
  The main resonance  peak power (for $p=1$ and $q=1$) is
the strongest among all the harmonics because the peak power
diminishes very quickly with the increase of $p$ and $q$ (LL). The
maximum of the main resonance amplitude for the nonlinear
oscillation is also not precisely at the eigenfrequency $\nu_0$.
In the nonlinear case the systematic and random shifts are (in addition to the damping)
affected by the amplitude of the driving force and the relative
weights (coefficients)  $\alpha_{k,l}$ of the nonlinear terms  of equations (1-2), 
(see LL for details). 

Recently, Abramovicz  et al. (2002)  drew attention of the community to the possible presense 
of $\nu_1/\nu_2=2:3$ resonance for two kHz frequencies.  A similar effect was also pointed out  
earlier by Abramovicz \& Klu\`zniak 2001; Remillard et al. (2002) for BH sources.
Possibly, it is not  by chance these kHz frequencies are detected in the observations because in 
anharmonic oscillator $\nu_1$ and $\nu_2$ would excite each other (if they are  related through this $2:3$ ratio,
see above). 
The oscillations with the  frequency $\nu_1$ ($\gamma$) would be a driving force for the oscillations with 
the eigenfrequency $\nu_2$ ($\nu_0$)  and  vice versa.  Abramovicz  et al. (2002) argue that it is hard to escape the
conclusion that a resonance is responsible for distribution of frequency ratios for Sco X-1 and 
other kHz QPO sources. But a 2:3 ratio is a natural consequence of the TL model where $\nu_2\approx\nu_h$ and
$\nu_1\approx \nu_{\rm K}$. In fact, $\nu_{\rm K }/\nu_{h}=[1+(2\nu_{mg}/\nu_{\rm K})^2]^{-1/2}$. 
The magnetospheric rotational frequency $\nu_{mg}$ should  very close to the NS spin frequency $\nu_{NS}= 
300-400$ Hz (at least within 15 \%).  But $2\nu_{mg}\sim2\nu_{NS}$ are very close the observed lower kHz frequency
$\nu_1$ ($\approx\nu_{\rm K}$) and thus  the ratio of $\nu_{\rm K}/\nu_{h}\approx 2^{-1/2}\approx 2:3$. 

The hybrid frequency $\nu_h$, introduced and derived in OT99, is a common feature of oscillations in a rotational
frame of reference under the influence of a gravitational force. When a gas (or fluid) has a nonuniform distribution 
of mass density, it may  exhibit the Rayleigh-Taylor instability. Hide (1956) and Chandrasekhar (1961) presented the
dispersion relation for a fluid rotating with angular velocity $\Omega$ ($\nu_r=\Omega/2\pi$)
$\nu_h^2[1-4\nu_r^2/\nu_h^2]^{1/2}= \omega_0^2$. It is easy to show that $\nu_0$ is very close to the Kepler
frequency and thus one can find that $\nu_h$, as a root of the Hide-Chandrasekhar relation, is very close to that obtained
by OT99. The detailed analysis of the Rayleigh-Taylor instability and oscillations 
for various density distributions, configurations and boundary conditions is given in Titarchuk (2002). 


\section{Results and Discussion}

It was shown in the previous section
that observed QPO frequencies $\nu_{HBO}$,  $\nu_1$ and $\nu_2$ could be treated 
as the resonance frequencies related to the eigenfrequencies of the linear damped 
oscillators.
One  can restore a particular eigenfrequency $\nu_0$, its error bar
$\Delta\nu_0$ using formulas (6)-(7) and 
 values of $\nu_r$, $\Delta\nu_r$, a QPO halfwidth, $\Delta\nu_{HW}=\lambda$.  
Thus for each set of $\nu_{HBO}\pm\Delta\nu_{HBO}$, $\nu_1\pm\Delta\nu_1$ and 
$\nu_2\pm\Delta\nu_2$ we obtain $\nu_L\pm\Delta\nu_L$, $\nu_{\rm K}\pm\Delta\nu_{\rm K}$ and 
$\nu_h\pm\Delta\nu_h$ as restored eigenfrequencies. 
According to formula (10) in OT99 $\nu_{mg}=(\nu_h^2-\nu_{\rm K}^2)^{1/2}/2$  is the
magnetospheric rotational frequency, which, for highly conductive plasma in the case of
corotation, should be approximately constant $\nu_{mg}=\nu_{mg,0}$.
In Figures 1 I present   $\nu_{mg}$ versus $\nu_{\rm K}$ for four Z-sources: 
Sco X-1, GX 340+0, GX 5-1, GX 17+2. Indeed one may notice that $\nu_{mg}=\nu_{mg,0}=const$
 is consistent for two Z-sources, Sco X-1 and GX 17+2. If the magnetosphere corotates with the neutron star
(solid-body) rotation, then this procedure of $\nu_{mg}-$extraction  determines the spin
rotation of the star.
For the other $Z-$sources,  GX 340+0, GX 5-1 one can find the features 
of the differential rotation which depend on the magnetic field ${\bf B}$ (see OT99).
These features of the differential rotation can  also be  identified for atoll sources 4U 1728-34 and 4U 0614+09.
The ''real`` errors bars $\Delta\nu_{mg}=(\nu_h\Delta\nu_h+ 
\nu_{\rm K}\Delta\nu_{\rm K})/4\nu_{mg}$ of $\nu_{mg}$  become larger due to these corrections
and consequently it is very difficult to identify significant deviations of the rotation of the 
magnetosphere  from the solid-body rotation (at least for two analyzed Z-sources).
The $\delta-$value  is calculated using formula (1) in TO01,
\begin{equation}
\delta=\arcsin{[(\nu_h^2-\nu_{\rm K}^2)^{-1/2}(\nu_L\nu_h/\nu_{\rm K})]}.
\end{equation}
The errors of $\delta$ is evaluated using the differential  
\begin{equation}
\Delta\delta=\delta[d\nu_L/\nu_L+d\nu_h/\nu_h+d\nu_{\rm K}/\nu_{\rm K}+
(\nu_hd\nu_h+\nu_{\rm K}d\nu_{\rm K})/4\nu^2_{mg}]. 
\end{equation}
In Figure~2 the dependence of 
the angle $\delta$ between a normal to the disk and a magnetospheric axis 
on $\nu_{\rm K}$ is shown for the four Z-sources
 The model invariants should be  the same, independent of the
eigenfrequency branch. Indeed, for each of these Z-sources
the $\delta-$values show little variation with $\nu_{\rm K}$, $\nu_h$, and $\nu_L$.
within the inferred error bars. The reduced $\chi^2_{red}$ for the best-fit $\delta$ is always
very close to 1.
Jonker et al (2002) presented the analysis of the $\delta-invariant$
of the TLM. They found that $\delta=6^0.3\pm 0\fdg1$ and
$\delta=6\fdg1\pm 0\fdg2$ for GX 5-1 and GX 17+2 respectively. But
they pointed out the very large $\chi^2_{red}=95.2/9$ and
$\chi^2_{red}=377/9$ for GX 5-1 and GX 17+2 respectively.  Their
$\chi^2$ calculations include the statistical error only but do not
account for the systematic resonance shift $\eta$ and for the random
resonance error $\Delta\nu_0$. It can be concluded that their 
$\chi^2$ for the $\delta-$mean
are significantly reduced when this effect is taken into account.  
For atoll sources in general it is not clear what kind of low frequencies can be used
for evaluation of $\delta$. For example, $\nu_{VLF}$ presented by van Straaten et al.  (2002)
for 4U 1728-34 have much stronger dependence on $\nu_{\rm 1}$ than those in 4U 0614+09. They
are rather subharmonics of $\nu_{VLF}$ (than $\nu_{HBO}$)
which are interpreted by Titarchuk, Bradshaw \& Wood
(2001) as the magnetoacoustic oscillation frequencies (see also Ford \& van der Klis 1998). 

\section{Conclusions} 
In this {\it Letter}, I present the  consequences and predictions of the TLM as an oscillatory model 
for the QPO observations. It is fair to ask what kind of  arguments one can put forth
for the resonance effect in QPO phenomena in general.  I argue that they  are (i)
the Lorentzian shape of the QPO feature (e.g. van der Klis 2000), (ii) $1/\omega$ resonance dependence 
of the rms amplitude for lower kHz QPO found by Titarchuk, Cui \& Wood (2002) using Jonker's et al. (2000b) data;
(iii) observational evidence for 2:3 resonance in the ``kHz'' QPO sources (Abramovicz \& Klu\`zniak 2001; 
Abramowicz et al. 2002), (iv) harmonics and sidebands of low kHz QPOs (Jonker et al. 2000b).  
The whole issue of the observational evidence for the resonance effect in the QPO phenomena is beyond 
the scope of this short {\it Letter}.

Using the previously cited thorough analysis of  RXTE QPO data for Sco X-1, 
Sco X-1, GX 340+0, GX 5-1, GX 17+2  and two atoll sources:   
4U 1728-34, 4U 0614+09 I a detailed investigation of the accuracy of 
the QPO frequency determination in the framework  of a weakly  nonlinear oscillator is offered.
The TLM  as an example of a weakly nonlinear oscillator  is used to  extract the model 
parameters and to test the model.    The TLM 
predicts the existence the  invariance of the $\delta$.
It is established that: (1) The errors of the eigenfrequency extraction
are significantly affected  by the errors of the QPO halfwidth.
(2)  The inferred $\delta-$values are consistent with 
constants for the four  Z-sources and 4U 0614+19 where kHz and HBO 
frequencies have been detected and  correctly identified.
I also put forth  arguments to explain the QPO phenomena 
as a result of the resonance effect in anharmonic oscillatory systems.
  

L.T. acknowledges  fruitful discussions with Vladimir Krasnopolsky,
Chris Shrader, Paul Ray and with referee.
In particular, I am grateful to Charlie Bradshaw, Michiel van der Klis, Peter Jonker and 
Mariano Mendez for the data  which enable me to make detailed comparisons of the model
predictions to the observations.

\begin{figure}
\includegraphics[width=6in,angle=0]{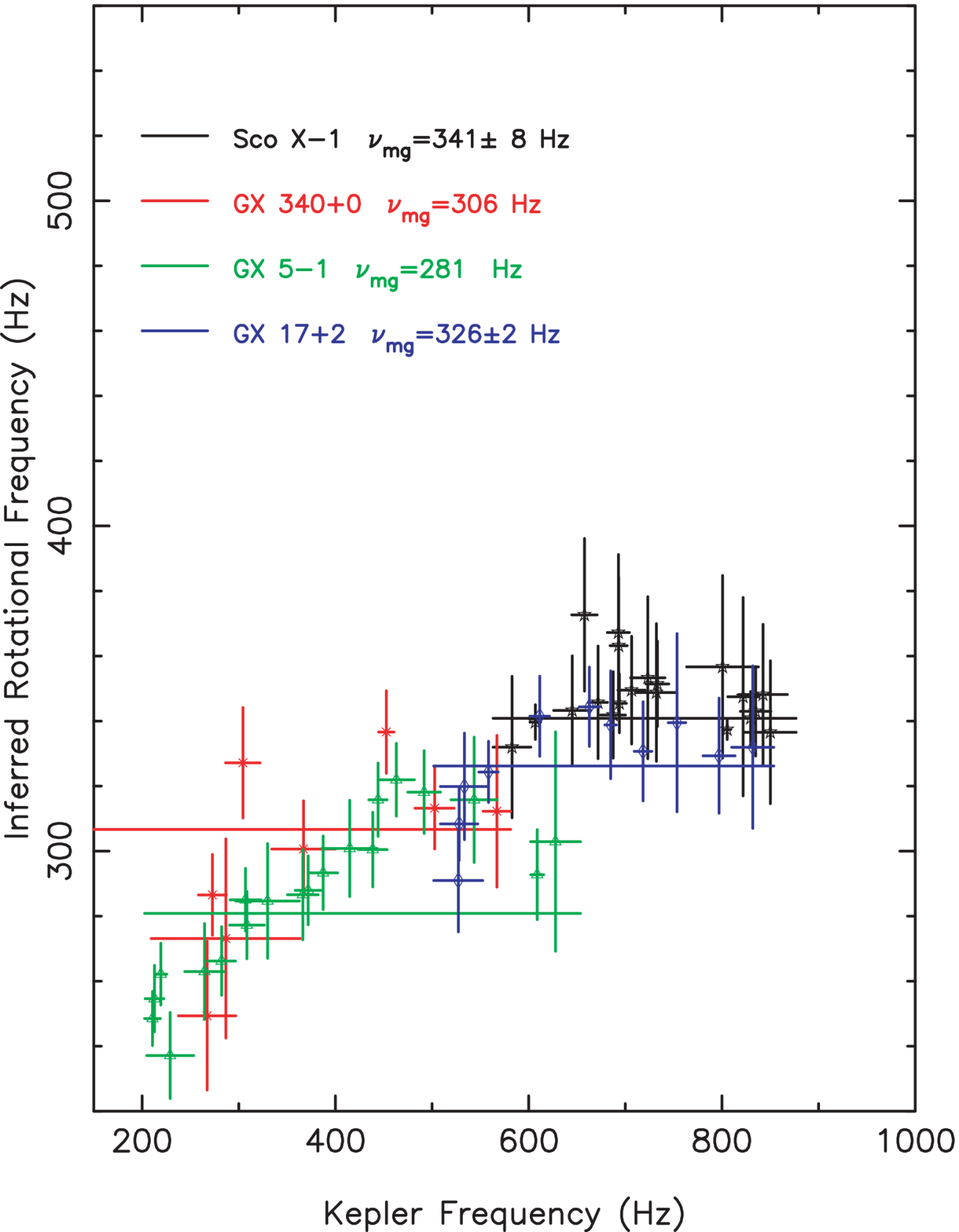}
\caption{\label{prof1}
Inferred rotational frequencies of NS magnetosphere as a function
of the Kepler frequency
 for a number of Z sources: Sco X-1 (black), GX 340+0 (red), GX 5-1 
 (green), GX 17+2 (blue). Constant rotational frequencies (possibly NS spin)
 for Sco X-1 and GX 17+2 (solid line) are consistent with the data. The  magnetospheric  rotational 
 profiles for GX 340+0 (red), GX 5-1  is inconsistent with  constant. The mean values of the rotational
 frequencies for these sources are given without error bars.
 }
\end{figure}


\begin{figure}
\includegraphics[width=6in,angle=0]{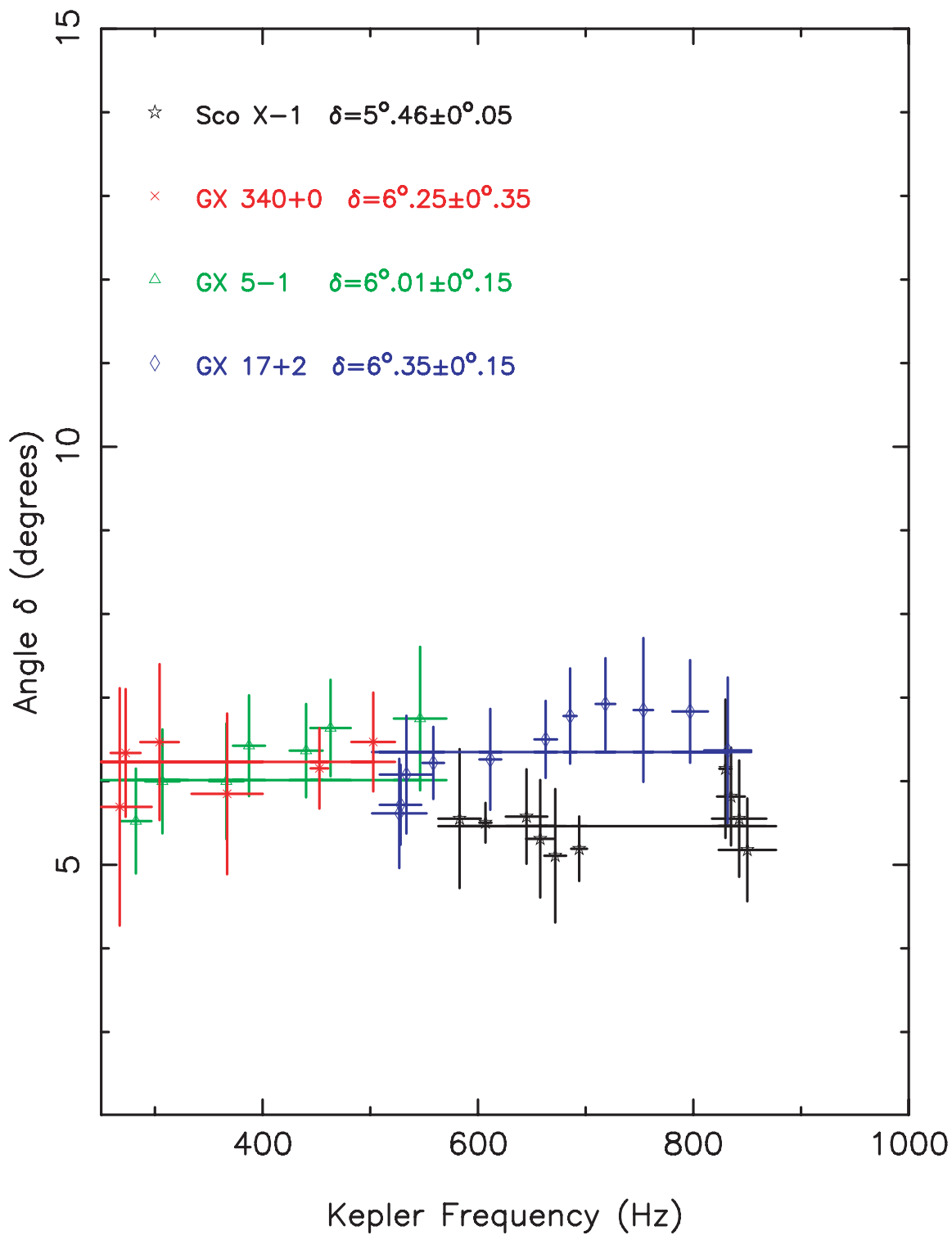}
\caption{\label{delta1}
Inferred angles $\delta$ between rotational 
  configuration above the disk (possibly NS magnetosphere) 
and   the normal to the plane of Keplerian oscillations as a function of 
  the Kepler frequency for Z-sources: 
   Sco X-1 (black), GX 340+0 (red), GX 5-1 
 (green), GX 17+2 (blue). Constant $\delta-$ value
 is consistent with the data. The $\delta-$angle (eq.[8]) is calculated using
 the lower and higher kHz peaks  $\nu_1$, $\nu_2$ and HBO frequencies $\nu_{HBO}$
(see text).
 }
\end{figure}


 

\clearpage

\end{document}